\documentclass{Interspeech2024}
\usepackage{amsmath}
\usepackage{multirow}
\usepackage{hyperref}
\usepackage{caption}
\usepackage{subcaption}
\usepackage{graphicx}%
\newsavebox{\measurebox}
\usepackage{array}
\usepackage{lipsum,capt-of,graphicx}
\usepackage[margin=1in]{geometry}
\usepackage{nicefrac, xfrac}
\usepackage{slashbox}
\usepackage[font=scriptsize]{caption}

\interspeechcameraready

\title{ROAR: Reinforcing Original to Augmented Data Ratio Dynamics for Wav2Vec2.0 Based ASR}

\name[affiliation={1}]{Vishwanath Pratap}{Singh}
\name[affiliation={1}]{Federico}{Malato}
\name[affiliation={1}]{Ville}{Hautamäki}
\name[affiliation={2,3}]{Md.}{Sahidullah}
\name[affiliation={1}]{Tomi}{Kinnunen}

\address{
  $^1$University of Eastern Finland, Finland\\
  $^2$Institute for Advancing Intelligence, TCG CREST, India \\
  $^3$Academy of Scientific and Innovative Research (AcSIR), India} 
\email{\{vsingh, federico.malato,tomi.kinnunen\}@uef.fi,\{ville,sahidullahmd\}@gmail.com}


\begin{document}

\maketitle
\begin{abstract}
While automatic speech recognition (ASR) greatly benefits from data augmentation, the augmentation recipes themselves tend to be heuristic. In this paper, we address one of the heuristic approach associated with balancing the right amount of augmented data in ASR training by introducing a reinforcement learning (RL) based dynamic adjustment of original-to-augmented data ratio (OAR). 
Unlike the fixed OAR approach in conventional data augmentation, our proposed method employs a deep Q-network (DQN) as the RL mechanism to learn the optimal dynamics of OAR throughout the wav2vec2.0 based ASR training. 
We conduct experiments using the LibriSpeech dataset with varying amounts of training data, specifically, the 10Min, 1H, 10H, and 100H splits to evaluate the efficacy of the proposed method under different data conditions. Our proposed method, on average, achieves a relative improvement of 4.96\% over the open-source wav2vec2.0 base model on standard LibriSpeech test sets. 

\end{abstract}

\vspace{-0.2 cm}
\section{Introduction}
\emph{Data augmentation} has emerged as a common strategy for model generalization and for increasing the quantity of training data to train the \emph{automatic speech recognition} (ASR) systems~\cite{aug_survey_asr}. Beyond merely increasing quantity, data augmentation also introduces diversity into the training dataset, thereby reducing the risk of overfitting~\cite{specaug,dl_aug}. Consequently, data augmentation has been extensively used for improving ASR systems in application-agnostic~\cite{povey_aug}, low-resource~\cite{low_aug,meng2021mixspeech}, multi-lingual~\cite{multi_aug}, and children's~\cite{child_aug} speech recognition scenarios.

While the advantages of data augmentation are apparent, a methodological challenge lies in selecting the optimal data augmentation methods, including their hyperparameters (such as signal-to-noise ratio (SNR) or speed modification factor), order (e.g., noise followed by speed modification, or noise followed by room impulse response (RIR)), and the volume of augmented data. In most of the previous studies~\cite{povey_aug, singhchildaugment}, the authors typically rely on heuristic ideas for choosing the augmentation methods, associated hyperparameters, and the amount of augmented data. Those are not necessarily optimum for different datasets and tasks; and choosing the right configuration remains an open challenge~\cite{lam2023make}.

In this paper, we tackle the aforementioned challenge by exploring automatic methods for setting specific data augmentation hyperparameters during training. In particular, we address automatic adjustment of the proportion by which the original and augmented data get mixed during different stages of training. For easier reference, we denote this quantity as the \emph{original-to-augmented data ratio} (OAR). 
Previous literature~\cite{povey_aug,snr_learn} highlights that exceeding the amount of augmented data from certain methods beyond a certain limit can lead to a degradation in ASR performance. On the other hand, if the augmented data ratio is kept too small, it might not lead to improvements because the limited input size may not significantly impact the ASR learning process. Thus, balancing this ratio emerges as a crucial factor in training robust and accurate ASR systems.

\begin{figure}[t]
\begin{minipage}[b]{1.0\linewidth}
  \centering
  \centerline{\includegraphics[width=5cm, height=14cm, keepaspectratio]{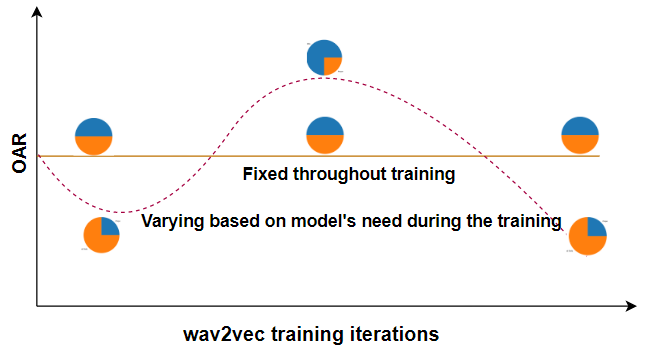}}
\end{minipage}
\footnotesize
\vspace{-0.5cm}
\caption{Two possible scenarios on dynamics of original-to-augmented data ratio (OAR) throughout the wav2vec2.0 training. The solid line (along with the fixed proportion in pie-chart) indicates fixed OAR used throughout the training. The dashed line (along with varying proportions in pie-chart) indicates our proposal, where OAR is allowed to vary dynamically throughout training.} 
\label{fig01a}
\vspace{-0.5 cm}
\end{figure}

\begin{figure*}[t]
\begin{minipage}[b]{1.0\linewidth}
  \centering
  \centerline{\includegraphics[width=9.75cm, keepaspectratio]{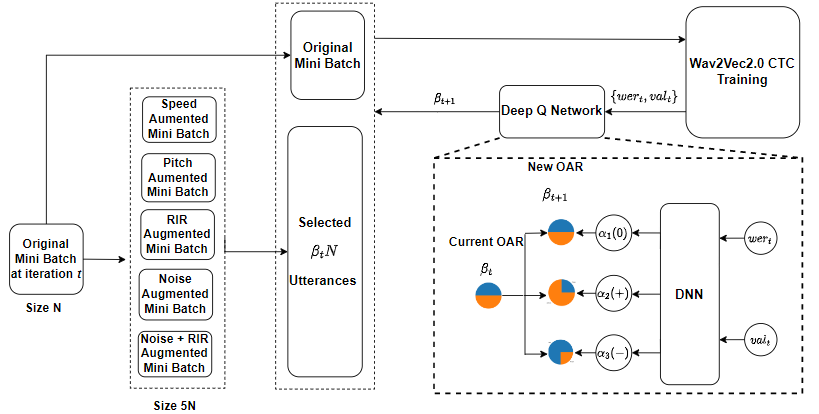}}
\end{minipage}
\vspace{-0.6cm}
\caption{Proposed ROAR based wav2vec2.0 CTC training pipeline, and illustration of Deep Q-network. $\beta_t$ indicates the original to augmented data ratio (OAR) at $t^{th}$ iteration of training, $val_t$ and $wer_t$ is validation loss and validation WER after $t^{th}$ iteration of wav2vec2.0 CTC training. $\alpha_1(0)$,  $\alpha_2(+)$, and $\alpha_3(-)$ indicate the actions corresponding to no change (i.e. null action), increase, and decrease in OAR, respectively.} 
\label{fig02}
\vspace{-0.65 cm}
\end{figure*}

Another fundamental question concerns the \emph{temporal dynamics} of the OAR throughout training. As illustrated in Fig. \ref{fig01a}, should this ratio remain constant---or would model training benefit from dynamic adjustments? We assume so, simply because the model's behavior evolves over the course of training. At the outset, when the model has not yet encountered substantial amounts of data, a lower OAR might be preferable, so as to better initialization of model parameters at the beginning of the training. However, as the model matures and starts adapting to the training set, a higher ratio may be beneficial for avoiding overfitting. In this paper, we hypothesize through experimentation that the dynamic adjustment of OAR is indeed beneficial in model training. 

Precisely, to address these concerns associated with dynamic adjustment of OAR, we propose \textbf{ROAR}, a novel approach that leverages \emph{reinforcement learning} (RL), so-called the specific RL approach, \emph{deep Q-network} (DQN)~\cite{dqn1}. The choice of RL as our workhorse is rooted in its capacity to address the intricate decision-making dynamics. We also chose DQN for its sample efficiency and ensured convergence properties~\cite{dqn1}. Sample efficiency is an important consideration as often ASR models are trained for a fixed number of iterations, which limits the number of samples available from the wav2vec2.0 training environment for DQN training. Moreover, our ASR training environment dynamics benefit from exploration strategies such as $\epsilon$-greedy~\cite{sutton2018reinforcement}. Our main contribution lies in proposing the DQN-based dynamic adjustment of OAR throughout the ASR training based on the ASR model's need for augmented data. 


\section{Related Work}
\vspace{-0.1cm}
Previously, automatic learning of augmentation policies using population based training (PBT) has been explored for automatic speech recognition (ASR) tasks \cite{pbt_audio}. However, to the best of our knowledge, RL-aided data augmentation has not been explored in the speech processing domain but considerable research has been conducted in the field of image processing~\cite{autoaug,autoaug-fast,tian2020improving} and natural language processing (NLP)~\cite{autoaug-text}. Specifically, authors in~\cite{autoaug, autoaug-fast} model the 3-dimensional augmentation parameters namely, the magnitude by which the augmentation is to be applied (such as \emph{signal-to-noise ratio} (SNR)), the augmentation policy such as rotation followed by translation, or the translation followed by rotation, and augmentation probability.

However, the idea introduced by us in this paper is different from those studied in the above image-processing and NLP domain. Firstly, we focus entirely on modeling the amount of augmented data through the OAR ($\beta$) as shown in Fig.~\ref{fig02}. Secondly, our RL training approach is different from those proposed in~\cite{autoaug, autoaug-fast}. Specifically, authors in~\cite{autoaug} use the neural architecture search with RL~\cite{neuralsearch}. On the other hand, we use a DQN-based RL strategy by deriving the reward for every $K^{th}$ iteration during the wav2vec2.0 CTC fine-tuning~\cite{wav2vec} which reduces the training time substantially. Thirdly, the motivation behind the augmentation in image processing is to create visual diversity by rotation, translation, and blurring of original images. Similarly, diversifying contextual diversity of particular words remains the key to NLP augmentations. While, in the case of speech, the augmentation methods revolve around the speaker, speaking style, and surrounding diversity. Hence, exploring the RL-aided augmentation for speech itself is one of the novel aspects of this paper.

\begin{figure*}[h]
\centering
  \begin{subfigure}{.32\textwidth}
    \includegraphics[width=.85\linewidth]{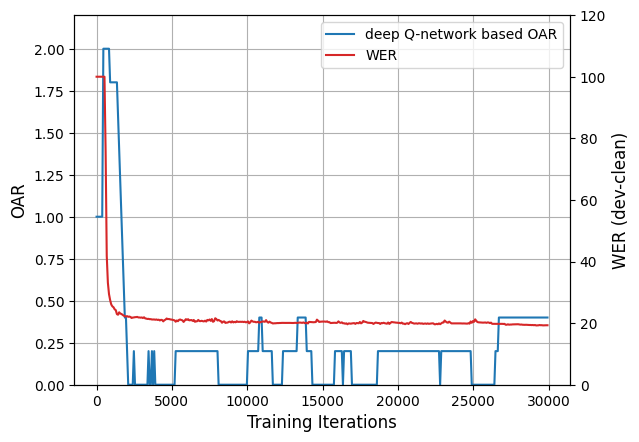}
    \caption{Episode-1: OAR dynamics} \label{fig03a}
  \end{subfigure}%
  \begin{subfigure}{.32\textwidth}
    \includegraphics[width=.85\linewidth]{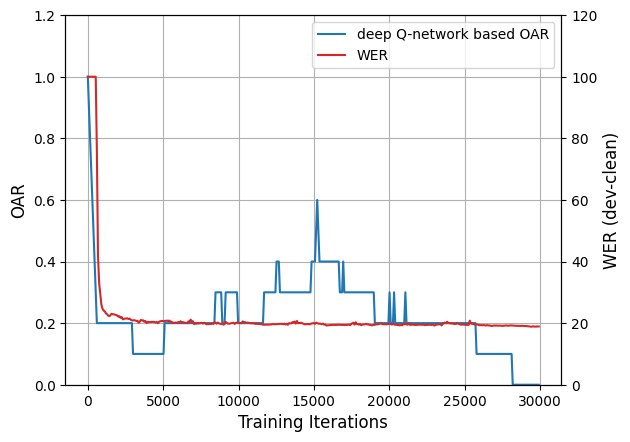}
    \caption{Episode-2: OAR dynamics} \label{fig03b}
  \end{subfigure}%
   \begin{subfigure}{.32\textwidth}
    \includegraphics[width=.85\linewidth]{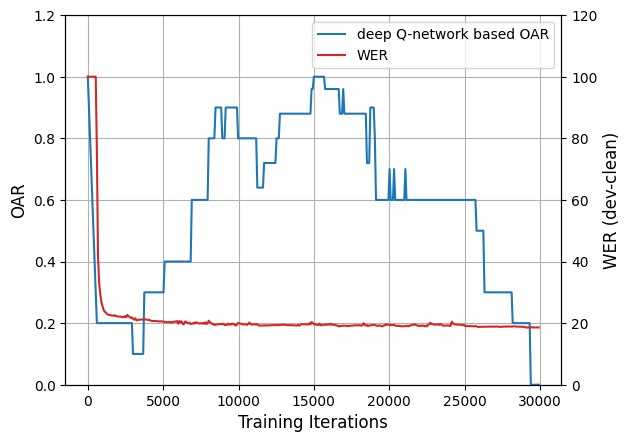}
    \caption{Episode-3: OAR dynamics} \label{fig03c}
  \end{subfigure}
  \vspace{-0.4 cm}
\caption{Depiction Deep Q-network based original-to-augmented data ratio (OAR) dynamics throughout the wav2vec2.0 training on Librispeech 1H training split and visualization of the evolution OAR across various episodes.} \label{fig03}
\vspace{-0.65 cm}
\end{figure*}
\vspace{-0.3cm}
\section{Proposed ROAR Method}
\vspace{-0.1cm}
\subsection{Background on DQN}
\vspace{-0.1 cm}
Deep Q-networks (DQN)~\cite{dqn1}, developed as a combination of \emph{Q-learning}~\cite{qlearning} and deep neural networks~\cite{nn}, is designed to handle high-dimensional state spaces by approximating the optimal action-value function. The core idea involves training a neural network to predict the Q-values for each possible action in a given state, enabling the agent to make informed decisions. Q-values signify the predicted cumulative reward that an agent expects to obtain by taking a specific action in a particular state. DQN agent interacts with an environment through a sequence of observations, actions, and rewards while storing the past experiences in a replay buffer. Past records are then used to update the policy, increasing the probability of selecting an action that maximizes the reward. DNN approximated Q value is modeled using a state value function given by:
\vspace{-0.2 cm}
\begin{equation}
\vspace{-0.2 cm}
\footnotesize
Q^{*}(s, a) = \max\limits_{\pi} \mathbb{E} \left[\sum_{\tau=1}^{T}{\gamma^{\tau} r_{t + \tau}} | s_{t} = s, a_t = a \right]
\end{equation}
where $Q^{*}(s, a)$ is the DNN approximated Q-value for state $s$, and action $a$, achievable by a policy $\pi = P(a|s)$, $r_t$ is the reward at step $t$,  $\gamma$ is the discount factor for future rewards, $T$ is the time horizon over which the sum is computed. 

During the training phase, the DQN agent systematically navigates the environment through multiple iterations, commencing exploration from the initial states and persisting until a terminal state or predefined horizon is attained. These iterative traversals are commonly denoted as training episodes. Across these episodes, the DQN agent consistently refines its policy, aiming for optimal decision-making.
\vspace{-0.2 cm}
\subsection{Using DQN to Adjust OAR}
\vspace{-0.1 cm}
Our proposed method, \emph{ROAR}, models the wav2vec2.0 CTC training~\cite{wav2vec} environment using a 2-dimensional space comprising validation loss and validation WER. 
We model the action space as a 3-dimensional, discrete space regulating the value of the current original to augmented data ratio. We set action 1 to be the null action, while actions 2 and 3 correspond to an increase and decrease of 0.2 of the current value, respectively. 
At each timestep, our DQN agent greedily selects the best action. Then, new batches of augmented data are generated and used for training wav2vec2.0 CTC model for a fixed number of iterations. Finally, we generate the reward from the change in WER. Hence, our proposed method benefits from the non-differentiable objective such as WER. Further, this approach offers a nuanced perspective, allowing the model to adapt to its evolving learning needs, ultimately enhancing the performance and robustness of wav2vec2.0~\cite{wav2vec} based ASR system. We show our DQN training setup and the overall wav2vec2.0 training pipeline in Fig.~\ref{fig02}.

\vspace{-0.2 cm}
\section{Experimental Setup}
\vspace{-0.1 cm}
\subsection{Dataset}
\vspace{-0.1 cm}
We utilize the Librispeech corpus~\cite{libri} to validate our proposed ROAR method. In particular, we use the same four training subsets as in~\cite{wav2vec} consisting of, respectively, 10 minutes, 1 hour, 10 hours, and 100 hours of data. We evaluate the baselines and proposed models on standard \texttt{dev-clean}, \texttt{dev-other}, \texttt{test-clean}, and \texttt{test-other} evaluation sets. \texttt{Dev-clean} is used as a validation set for deriving the reward for RL-based training pipeline, and also selecting the best (in terms of validation WER) baselines and proposed model checkpoints. 
\vspace{-0.2 cm}
\subsection{ASR System}
\vspace{-0.1 cm}
We consider the frequently used open-source wav2vec2.0~\cite{wav2vec} based ASR system in our experiments. The wav2vec2.0 is available in two configurations, namely, \emph{Base} and \emph{Large}. We utilize the Base configuration for our experimentation which includes 12 transformer~\cite{vaswani} blocks, a model dimension of 768, an inner dimension of 3,072, and 8 attention heads, totaling 94 million parameters.

The wav2vec2.0 training comprises two stages. The initial stage involves pretraining, which is a contrastive loss~\cite{closs} based unsupervised training aimed at generating application-independent audio embeddings. Checkpoints available from this stage of training are referred to as \emph{self-supervised} checkpoints. In the second stage, referred to as the fine-tuning phase, an output layer is added, and the entire model is fine-tuned with CTC~\cite{ctc} loss. Checkpoints available from this stage of training are referred to as \emph{pre-trained} ASR. All publicly available wav2vec2.0~\cite{wav2vec} self-supervised checkpoints and pre-trained ASR are trained without augmentation. 

In our experiments, we forego the first stage training and instead utilize open-source self-supervised wav2vec2.0 base checkpoint \footnote{Available as of February 2024: https://dl.fbaipublicfiles.com\/fairseq\/wav2vec\/wav2vec\_small.pt} and finetune them using CTC loss~\cite{ctc}. We explore two scenarios in the second stage CTC training: in the first scenario, we randomly initialize the output layer on top of the self-supervised wav2vec2.0 checkpoint and train the model under different data conditions. In the second scenario, we utilize the open-source pre-trained ASR wav2vec2.0 model \footnote{Available as of February 2024: https://dl.fbaipublicfiles.com/fairseq/wav2vec/wav2vec\_small\_100h.pt}, already trained on 100 Hours LibriSpeech using CTC loss, and further train it on augmented data. 

The rationale behind the second scenario is rooted in the working mechanism of DQN. Our DQN derives reward from the change in WER, which is expected to decrease substantially at the beginning of training in the first scenario, as the weights are randomly initialized and WER is at its max. Hence, the reward is always positive for DQN at the beginning, even if it makes a few wrong decisions. On the other hand, in the second scenario, the open-source fine-tuned wav2vec2.0 is already optimal and only the right OAR (i.e. right DQN action) will lead to the decrease in WER. This ensures the reinforcement of DQN from the beginning of the training.

\begin{table*}[h!]
 \footnotesize
 \caption{Results (in terms of \%WER) with self-supervised wav2vec2.0 Base checkpoint trained on different amounts of labeled data scenarios. $OAR={0,1,2,3,4}$ indicates the fixed OAR throughout the wav2vec2.0 training. Results in boldface indicate the best baseline and proposed ROAR based results.}
 \label{tab:res1}
 \centering
 \vspace{-0.3 cm}

 \begin{tabular}{|c | c | c | c |c |c |} 
 \cline{3-6}
 \multicolumn{2}{ c| }{} &\multicolumn{4}{c|}{LibriSpeech Evaluation Splits} \\
\cline{1-6}

\multicolumn{1}{ |c| }{ Labeled Data} & \multicolumn{1}{ c| }{OAR} & \multicolumn{1}{ c| }{Dev-Clean} & \multicolumn{1}{ c| }{Dev-Other} & \multicolumn{1}{ c| }{Test-Clean} & \multicolumn{1}{ c| }{Test-Other}\\ 
 \cline{2-6}\hline
  \multirow{7}{*}{10 Minutes}& 0 (no augmentation) & 40.7& 48.8& 41.5 & 48.9\\ 
  &1 & 39.1 & 47.3 & 40.0& 48.0\\
  &2 &\textbf{39.0} & \textbf{47.3}& \textbf{39.8}&\textbf{47.5} \\
  &3 & 40.1 &  47.8 & 40.4 & 48.0\\
  &4 & 40.4  & 48.2& 40.8 & 48.7 \\
  &ROAR & \textbf{37.15}  & \textbf{45.9} & \textbf{38.11} & \textbf{46.0} \\
 \hline \hline
  \multirow{6}{*}{1 Hour}& 0 (no augmentation) & 19.5 & 29.5 & 20.2 & 20.4\\ 
  &1 & \textbf{19.3} &  29.2 &  \textbf{19.9} & \textbf{29.8} \\
  &2 & 19.6 &  \textbf{29.1} & 20.0 & 30.0 \\
  &3 &19.9 & 30.2 & 20.5 & 30.8 \\
  &4 & 20.2 & 30.8 & 21.0 & 31.3\\
  &ROAR & \textbf{18.6}& \textbf{28.4} &\textbf{19.1} & \textbf{28.9}\\
\hline \hline
\multirow{6}{*}{10 Hours}& 0 (no augmentation) & 9.9 &  19.3 & 10.1& 19.5 \\ 
  &1 & \textbf{9.5} &  \textbf{18.2} & \textbf{9.5} & \textbf{18.4} \\
  &2 & 9.6 & 18.5 & 9.7 & 18.5 \\
  &3 & 9.9 &  18.9 & 10.0 & 18.7\\
  &4 & 10.2 &  19.4 & 10.5 & 19.9 \\
  &ROAR & \textbf{9.1} & \textbf{17.7} & \textbf{9.3} & \textbf{17.7}\\
\hline \hline
\multirow{7}{*}{100 Hours}& open-source & 6.1 & 13.8 & 6.1& 13.5\\ 
  &0 (no augmentation) &6.0 &  14.1& 6.1& 13.9\\ 
  &1 & 5.8 & \textbf{13.7} & \textbf{5.9}& \textbf{13.5}\\
  &2 & \textbf{5.6} & 13.8 & 6.0 & 13.9\\
  &3 & 6.2 & 14.1 & 6.1& 14.0\\
  &4 & 6.1& 14.0 & 6.0 & 13.9\\
  &ROAR & \textbf{5.3} & \textbf{13.3} & \textbf{5.6} & \textbf{13.1} \\
\hline
 \end{tabular}
 \vspace{-0.2 cm}
\end{table*}
\begin{table*}[h!]
\footnotesize
 \caption{Results (in terms of \%WER) with pre-trained wav2vec2.0 Base model further trained on LibriSpeech  100 hours training split along with different augmentation strategies. The confidence interval is computed over the models obtained from 5 different runs. }
 \label{tab:res2}
 \centering
 \vspace{-0.35 cm}
 
 \begin{tabular}{|c | c | c | c |c |c |} 
 \cline{3-6}
 \multicolumn{2}{ c| }{} &\multicolumn{4}{c|}{LibriSpeech Evaluation Splits} \\
\cline{1-6}

\multicolumn{1}{ |c| }{ Labeled Data} & \multicolumn{1}{ c| }{OAR} & \multicolumn{1}{ c| }{Dev-Clean} & \multicolumn{1}{ c| }{Dev-Other} & \multicolumn{1}{ c| }{Test-Clean} & \multicolumn{1}{ c| }{Test-Other}\\ 
 \cline{2-6}\hline
\multirow{7}{*}{100 Hours}& open-source & 6.1  & 13.8 & 6.1 & 13.5 \\
&0 &$6.1 \pm 0.1$ & $13.7 \pm 0.1$ & $6.1 \pm 0.1$ &  $13.4 \pm 0.1$\\
  &1 &$6.1 \pm 0.1$ &  $13.4 \pm 0.2$ & $6.1 \pm 0.1$ & $13.2 \pm 0.2$\\
  &2 &$6.1 \pm 0.1$ &  $13.5 \pm 0.2$ & $6.0 \pm 0.1$ & $13.0 \pm 0.1$\\
  &3 &$6.0 \pm 0.1$ &  $13.4 \pm 0.3$ & $6.0 \pm 0.1$ & $12.9 \pm 0.2$\\
  &4 &$6.2 \pm 0.2$ &  $13.6 \pm 0.3$ & $6.1 \pm 0.2$ & $13.2 \pm 0.3$\\
  &ROAR & $5.8 \pm 0.2$ & $13.1 \pm 0.2$ & $5.9 \pm 0.1$& $12.6 \pm 0.2$ \\
\hline
 \end{tabular}
 \vspace{-0.4 cm}
\end{table*}
\vspace{-0.2 cm}

\subsection{Baseline}
\vspace{-0.1 cm}
Our baseline ASR systems include state-of-the-art wav2vec2.0 trained with CTC loss for different but fixed original to augmented data (OAR). OAR $0$  indicates the standard wav2vec2.0 CTC training where no augmentation is used. For reference, we also include the results with the open-source wav2vec2.0 model (wherever available). 

\vspace{-0.2 cm}
\subsection{Data Augmentation Methods}
\vspace{-0.1 cm}
In this study, we employ the following commonly used five data augmentation techniques to generate augmented data in experimentation:

\textbf{\emph{Noise}}: The addition of background noise to the audio data with signal-to-noise (SNR) sampled uniformly between $0–20$~dB, simulating real-world conditions~\cite{musan}. \textbf{\emph{RIR}}: Room impulse response-based augmentations that emulate different acoustic environments~\cite{rir}. \textbf{\emph{Noise and RIR}}: First Noise and then RIR are applied simultaneously on time domain speech signal. \textbf{\emph{SM}}: Speed modification with factor sampled uniformly between $0.9-1.1$~\cite{aug_survey_asr}. \textbf{\emph{PM}}: Pitch modification with factor sampled uniformly between $0.9-1.1$~\cite{aug_survey_asr}.
\vspace{-0.2cm}
\subsection{Deep Q-Network}
\vspace{-0.1cm}
Our DQN model configuration includes the most commonly used \emph{epsilon-greedy} Q-policy~\cite{qlearning} with a learning rate set to 0.001, a discount factor $\gamma$ of 0.99, warm-up steps 50, a replay buffer size of 10,000, and a batch size of 32 for experience replay. The neural network architecture consisted of two fully connected layers with 64 neurons each, employing rectified linear unit (ReLU) activation functions~\cite{relu}. As shown in Fig.~\ref{fig02}, input to the DQN is a 2-dimensional state modeled using validation loss and WER of wav2vec2.0 training, and output action space is 3-dimensional. 

\vspace{-0.2cm}
\section{Results and Discussion}
\vspace{-0.1cm}
\subsection{Baselines}
\vspace{-0.1cm}
We observe in Table~\ref{tab:res1} that for 10 minutes of fine-tuning data, the baseline trained with fixed OAR of $2$ outperforms the remaining baselines along with standard wav2vec2.0 trained with OAR of $0$ (i.e. without augmentation). This indicates the usefulness of augmentation in wav2vec2.0 training. Similarly, in the case of 1 hour of fine-tuning data, the baseline trained with fixed OAR of $1$ outperforms the remaining baselines on 3 out of 4 evaluation sets. This might indicate that a lesser amount of augmented data is required as the amount of training data increases. This phenomenon persists across the 10 hours and 100 hours models as well, where a fixed OAR of $1$ yields the optimal result. Moreover, we observe in Table~\ref{tab:res1} that our baseline system, trained with fixed OAR with LibriSpeech 100 hours, outperforms the corresponding open-source wav2vec2.0 pre-trained ASR on \texttt{dev-clean} and \texttt{test-clean} evaluation sets with on an average $5.5\%$ of relative improvement.

Further, observations of Table~\ref{tab:res2} reveal that baselines trained with a fixed OAR of $3$ outperform those trained with other OARs. This trend could be attributed to the initialization of wav2vec2.0 with a pre-trained ASR checkpoint, which has already undergone comprehensive training on the original LibriSpeech 100 hours dataset. Hence, more diversity in training data is expected by the model. 
\vspace{-0.2 cm}
\subsection{ROAR Based Wav2Vec2.0}
\vspace{-0.1 cm}
We observe in Table~\ref{tab:res1} that ROAR based model archives on an average improvement of 3.75\%, 3.26\%, 3.35\%, 4.07\% over best baselines (trained with fixed OAR) on 10 minutes, 1 hour, 10 hours, and 100 hours of LibriSpeech, respectively. Further, we observe in Table~\ref{tab:res2} that the ROAR based model, further trained on wav2vec2.0 pre-trained ASR, outperforms the open-source wav2vec2.0 pre-trained ASR with an average 4.96\% as well as the best baseline trained with fixed OAR with an average relative improvement of 2.4\% of relative improvement across standard LibriSpeech evaluation sets. This indicates the significance of the dynamic adjustment of OAR in wav2vec2.0-based ASR training.

Moreover, for stability analysis, we also present the results along with confidence interval, over 5 different runs, in Table \ref{tab:res2}.
\vspace{-0.2 cm}
\subsection{OAR Dynamics}
\vspace{-0.1 cm}
Deep Q-network based OAR dynamics and the evolution of OAR dynamics across further episodes in the context of wav2vec2.0 CTC training for LibriSpeech 1H are illustrated in Fig.~\ref{fig03}. We observe in Fig.~\ref{fig03a} that OAR fluctuates but remains very small throughout the wav2vec2.0 training. This behavior might be attributed to the random initialization of DQN weights at the beginning of the \emph{episode 1} and the limited interaction of DQN with the wav2vec2.0 training environment. As the training further proceeds, we observe in the OAR dynamics of \emph{episode 2} that the DQN is trying to keep the OAR small at the beginning and end of the wav2vec2.0 training and relatively higher OAR in the mid-iterations of training. This phenomenon is further reinforced in \emph{episode 3}, where OAR is once again small at the outset of training, increases in a stepwise fashion during mid-range iterations, remains elevated throughout this phase and eventually decreases to 0 in a stepwise manner at the end of training. These observations hypothesize our intuition depicted in \ref{fig01a} and claimed in Section 1.

\vspace{-0.1cm}
\section{Conclusion}
\vspace{-0.1cm}
This study demonstrates that dynamically adjusting the OAR using DQN provides advantages over the commonly employed fixed OAR approach for wav2vec2.0 based ASR. While the benefit of DQN-based OAR dynamics is evident, our study also introduces a few limitations to the proposed approach. First, the validation WER is expected to decrease at the beginning of ASR training, and hence DQN might not get reinforced very well at the beginning of training. Second, the DQN agents are initialized randomly at the beginning of episode 1, and hence a detailed stability analysis on OAR dynamics will be beneficial. 

While in this study, we focus entirely on optimizing the OAR dynamic through DQN, it can be extended by jointly optimizing other augmentation hyperparameters such as modification factors (e.g. signal-to-noise ratio, speed modification factor) along with OAR.
\vspace{-0.2 cm}
\section{Acknowledgment}
\vspace{-0.1cm}
This work was partially supported by Academy of Finland (Decision No. 349605, project ``SPEECHFAKES'')
\newpage

\bibliographystyle{IEEEtran}
\bibliography{mybib}
\end{document}